# Aortic frequency response determination via bioimpedance plethysmography

Roman Kusche, Arthur-Vincent Lindenberg, Sebastian Hauschild, and Martin Ryschka

*Abstract— Objective:* Arterial stiffness is an important marker to predict cardio vascular events. Common measurement techniques to determine the condition of the aorta are limited to the acquisition of the arterial pulse wave at the extremities. The goal of this work is to enable non-invasive measurements of the aortic pulse wave velocity, instead. An additional aim is to extract further information, related to the conditions of the aorta, from the pulse wave signal instead of only its velocity.
*Methods:* After discussing the problems of common pulse wave analysis procedures, an approach to determine the frequency response of the aorta is presented. Therefore, the aorta is modelled as an electrical equivalent circuit. To determine the specific numeric values of this system, a measurement approach is presented, which is based on non-invasive bioimpedance plethysmography measurements above the aortic arch and at the inguinal region. The conversion of the measurement results to the system parameters is realized by a digital algorithm, which is proposed in this work as well. To evaluate the approach, a study on three subjects is performed. *Results:* The measurement results demonstrate that the proposed approach yields realistic frequency responses. For better approximation of the aortic system function, more complex models are recommended to investigate in the future. Since this study is limited to three subjects without a ground truth, further measurements will be necessary. *Significance:* The proposed approach could solve problems of current methods to determine the condition of the aorta. Its application is non-invasive, harmless and easy to execute.

## I. Introduction

CARDIO vascular diseases (CVD), such as heart attacks or ischemic strokes, are the most often causes of deaths worldwide [1]. There is a strong relationship between the risk

Manuscript received xxx xx, 201x; revised xxx xx, 201x and xxx xx, 201x; accepted xxx xx, 201x. Date of publication xxx xx, 201x; date of current version xxx xx, 201x. This research was funded by the German Federal Ministry of Education and Research (BMBF, FKZ 13EZ1140A/B).
R. Kusche, A.-V. Lindenberg, S. Hauschild and M. Ryschka are with the Laboratory of Medical Electronics (LME), Luebeck University of Applied Sciences, 23562 Luebeck, Germany. (email: roman.kusche@th-luebeck.de; martin.ryschka@th-luebeck.de).
Color versions of one or more of the figures in this paper are available online at http://ieeexplore.ieee.org.
Digital Object Identifier xx.xxxxx/xxxxxx

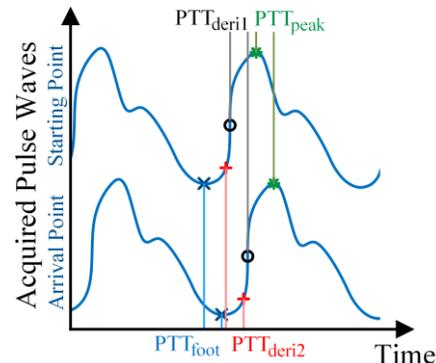

Fig. 1. Common determination of the pulse transit time by acquiring the pulse wave at two different points. The four typical characteristic signal points (foot, first derivation, second derivation, peak), which are commonly used for PTT measurements, are marked.

of a cardio vascular disease and the arterial stiffness. Especially, the stiffness of the aorta, the dominant fluid reservoir, which dampens the pressure pulses, is a powerful indicator and therefore of interest to be measured [2]. Since it is difficult to determine the aortic mechanical characteristic non-invasively, several alternative measurement procedures and parameters have been used in the past [3-6].

The most common method to gather information about the aorta is the analysis of the pressure pulse wave, which is generated by the pumping of the heart [7-9]. Its morphology, as well as its velocity within the arteries contain useful information about the blood vessels' conditions. The resulting parameters are the augmentation index (AIx) and the pulse wave velocity (PWV), which are described more detailed in the materials and methods section.

A major problem of the pulse wave analysis is the difficulty to detect the pulse wave directly at the aorta. Instead of this, it became very popular to acquire the pulse wave at one or two peripheral arteries [4, 6, 10, 11]. Afterwards, complex models and algorithms are used to estimate the actual wave morphologies within the aorta. It is conceivable, that modelling the system, which shall actually be determined can provoke significant errors.

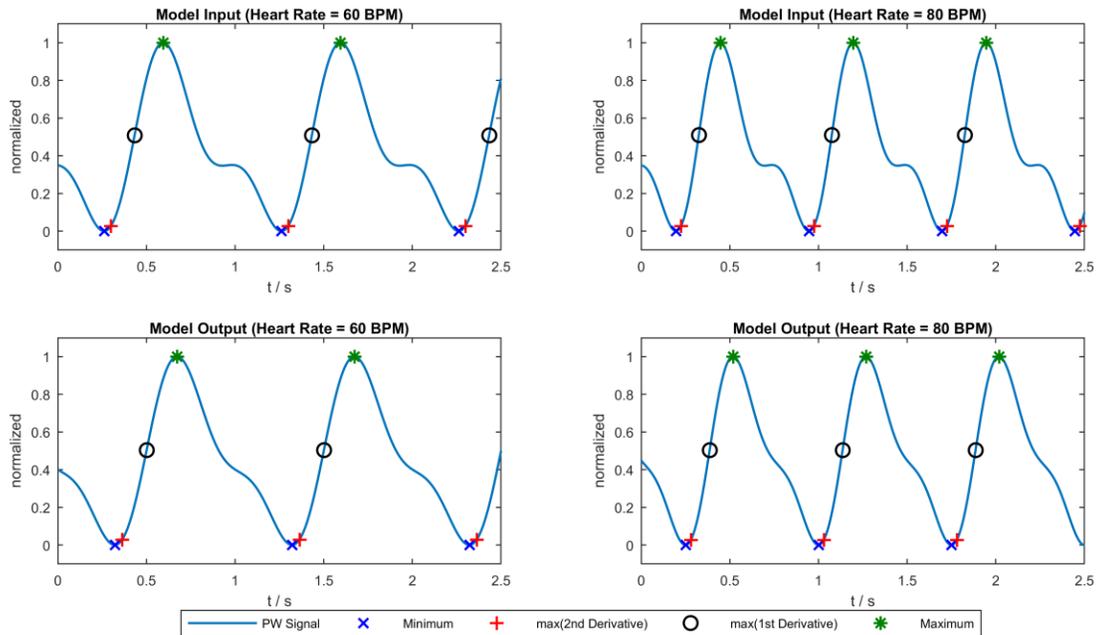

Fig. 2. Simulated pulse waves, representing heart rates of 60 BPM and 80 BPM (upper plots), which are filtered by a first order low-pass filter (lower plots) to simply emulate the mechanical behavior of the aorta. The characteristic pulse wave points for a typical PTT determination of each signal are marked.

TABLE I
SIMULATION RESULTS OF THE PULSE TRANSIT TIME FOR THE FOUR COMMON DETECTION POINTS

| Simulated Heart Rate | $PTT_{foot}$ / ms | $PTT_{deri2}$ / ms | $PTT_{deri1}$ / ms | $PTT_{peak}$ / ms |
|---|---|---|---|---|
| 60 BPM | 61 | 64 | 68 | 79 |
| 80 BPM | 54 | 56 | 61 | 72 |

This work examines the problems of common pulse wave analysis procedures and proposes the determination of the aortic frequency response instead. Therefore, a bioimpedance based measurement setup is suggested, which acquires the pulse waves at the aortic arch and at the end of the aorta simultaneously. Afterwards, an electrical equivalent circuit to model the mechanical characteristics of the aorta is proposed. To extract the model parameters from the measured pulse wave signals, a digital algorithm is suggested. Finally, the bioimpedance based measurement setup is used to acquire aortic pulse wave signals from three subjects. The gathered signals are processed with the suggested signal processing algorithm to estimate the corresponding aortic frequency responses.

## II. MATERIALS AND METHODS

### A. Conventional pulse wave analysis

*1) Pulse wave velocity:*
The measurement of the PWV is based on the fact that the velocity of a fluid within an elastic vessel depends on its stiffness [12]. A simplified equation to describe this physical relationship is the Moens-Korteweg equation [13], shown in equation 1, which is often used in the field of PWV researches. In this equation, E represents the elastic modulus and h represents the arterial wall thickness. Additionally, the PWV is influenced by the vessel radius r and the blood density ρ. Typically, the PWV varies between 5 m/s for young and healthy subjects and 15 m/s for subjects with a high risk of an upcoming cardio vascular event [14].

$$PWV = \sqrt{\frac{E \cdot h}{2 \cdot r \cdot \rho}} \quad (1)$$

There are two common kinds of techniques to measure the aortic PWV. The first method is to detect the pulse wave at just one point, commonly at the extremities, with a pressure sensor which is positioned above an artery, e. g. the arteria radialis [6]. The morphology of the acquired pulse wave in combination with several additional subject information, for instance sex, age, weight and height, is then used to estimate the actual PWV within the aorta. This method is primary based on the assumption that specific reflections occur inside the aorta, which influence the pulse wave morphology.

The second method is more straightforward. An additional sensor is used to realize a pulse transit time (PTT) measurement. Obviously, it would be reasonable to measure the PTT between the beginning and the end of the aorta. Since it is difficult to access these points non-invasively, some works have been published, in which the pulse waves are acquired at the extremities instead [6].
Combining the knowledge of the measured PTT and of the distance Δx between both sensors yields to the mean PWV within this vessel, as shown in equation 2.

$$PWV_{mean} = \frac{\Delta x}{PTT} \quad (2)$$

The parameter PWV itself, but also existing measurement setups are very error-prone [15]. Other researchers have shown that the ten-year cardio vascular disease risk is about 10 % when the PWV is about 8 m/s, but already double when the PWV is about 11 m/s [16]. Therefore, useful PWV measurements have to provide tolerances in the range of 1

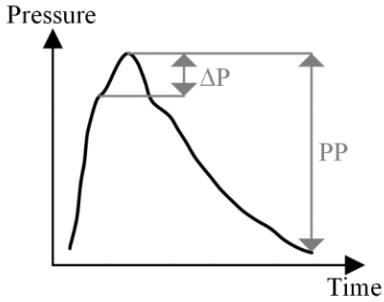

Fig. 3. Illustration of an arterial pulse wave and the determination of the pressure increase ΔP and the total pressure PP for calculating the augmentation index.

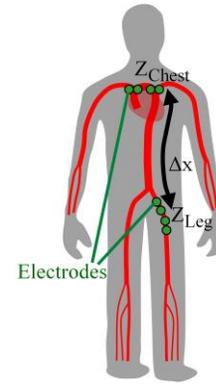

Fig. 4. Principle measurement setup to acquire the pulse wave at the beginning and at the end of the aorta via bioimpedance plethysmography.

m/s, respectively about 10 %.

The first problem regarding the determination of PWV is the Moens-Korteweg equation. This approximation was originally not developed for arteries, but rather to describe the velocity of waves in elastic tubes or hoses [13]. Therefore, applying it to arteries, presumes the aorta to be a straight ideal hose and the arterial wall to be isotropic. This is obviously not the case and thus a compromise. In equation 1 it can also be seen, that not just the elasticity of the vessel influences the PWV, but also the wall thickness h and the vessel radius r. Since these values are unknown and can usually just be estimated, the PWV cannot be used as a synonym for arterial stiffness, but indicator at best.

The second problem is the measurement instrumentation. As described before, some devices estimate the aortic PWV by detecting the pulse wave at just one peripheral point. Therefore, the precise length of the aorta has to be estimated. It is conceivable, that this value is not same for all subjects and the deviations have a linear impact on the measurement results, as it can be seen in equation 2.

Not only the single-point, but also the two-point measurement setup is problematic. The determination of the transit time between the pulse wave starting point and the arrival point is typically performed by detecting characteristic points of the pulse wave [17]. In figure 1, exemplary pulse wave signals are depicted and the most commonly used characteristic points are marked. From the implementation point of view, surely the detection of the pulse wave feet or peaks is the simplest. Some publications suggest the usage of the maximum values of the first (deri1) or the second (deri2) derivative as characteristic time points [18].

Unfortunately, this approach neglects the dampening characteristic of the vessel, which can be modelled very elementary as a simple first order low-pass filter with the time constant T [19]. The corresponding frequency response $H_{LP1}(j\omega)$ as a function of the angular frequency $\omega = 2\cdot\pi\cdot f$ is shown in equation 3.

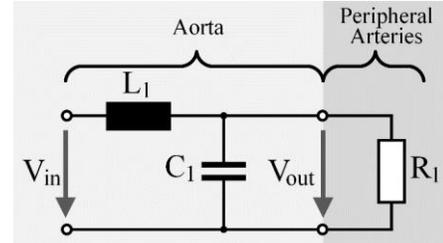

Fig. 5. Electrical equivalent circuit of the aorta.

$$H_{LP1}(j\omega) = \frac{1}{1+j\omega T} \quad (3)$$

The corresponding phase response φ(ω) of this model is shown in equation 4.

$$\phi(\omega) = -\arctan(\omega T) \quad (4)$$

Differentiating the phase response with respect to ω leads to the system's group delay $\tau_g(\omega)$ in equation 5, which describes the occurring delay of each frequency component. This dispersion is significant, but in other approaches often neglected.

$$\tau_g(\omega) = -\frac{d\phi(\omega)}{d\omega} = \frac{T}{1+\omega^2 T^2} \quad (5)$$

Since the pulse wave is not a sinusoidal signal, but contains several frequency components, this group delay has to be considered. The resulting different frequency dependent transit times of the pulse wave signal components yield a change of the signal morphology, when passing the blood vessels. This signal distortion can have a significant influence on the previously described characteristic points of a pulse wave.

To demonstrate the impact of this issue, two simple artificial pulse wave signals ($PW_{60BPM}$, $PW_{80BPM}$), representing two different heart rates of 60 beats per minute (BPM) and 80 BPM are generated, as described in equation 6 and 7.

$$PW_{60BPM}(t) = \sin(2\pi \cdot 1\,Hz \cdot t) + 0.5 \cdot \sin\left(2\pi \cdot 2\,Hz \cdot t - \frac{\pi}{6}\right) \quad (6)$$

$$PW_{80BPM}(t) = \sin\left(2\pi \cdot \frac{4}{3}\,Hz \cdot t\right) + 0.5 \cdot \sin\left(2\pi \cdot \frac{8}{3}\,Hz \cdot t - \frac{\pi}{6}\right) \quad (7)$$

TABLE II
RELATIONSHIP BETWEEN THE ELECTRICAL EQUIVALENT CIRCUIT AND THE MECHANICAL PARAMETERS

| Symbol | Electrical Parameter | Electrical Unit | Mechanical Parameter | Mechanical Unit |
|---|---|---|---|---|
| I | Current | A | Volumetric flow rate | $\frac{m^3}{s}$ |
| V | Voltage | V | Pressure | $Pa$ |
| Q | Charge | As | Blood volume | $m^3$ |
| C | Capacity | F | Ability to store blood volume | $\frac{m^3}{Pa}$ |
| L | Inductance | H | Mass inertia | $\frac{kg}{m^4}$ |

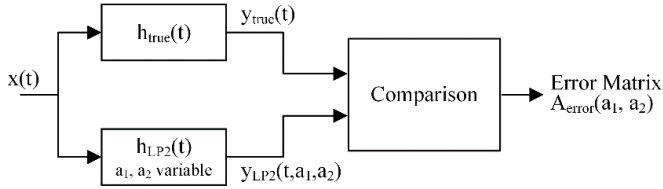

Fig. 6. Block diagram of the procedure to determine the optimum combination of the parameters $a_1$ and $a_2$.

Both signals are depicted in the upper plots of figure 2 with the corresponding characteristic points. To simply emulate the aorta, a first order low-pass filter with a cut-off frequency of $f_c=1.8$ Hz is applied to the signals and the resulting outputs are depicted in the lower plots of figure 2 with the corresponding characteristic points. The corresponding calculated PTT values are listed in table 1. It can be seen, that the calculated pulse transit times vary widely, depending on the used characteristic points to calculate them. In this example, the frequency dependent group delay leads for example to an increase of about 30 % of the calculated PTT values when using the maxima of the 80 BPM signal instead of the signal minima points. Additionally, it can be seen, that already a minor change of the heart rate from 60 BPM to 80 BPM causes significant measurement derivations, even when using the same characteristic signal points.
Even when this example is based on very simple assumptions and may not fit well to real pulse waves and the mechanical characteristics of the aorta, it demonstrates the sensitivity of this measurement method to unknown parameters.

*2) Augmentation Index:*

The determination of the AIx is a measurement method which is based on the acquisition of the pulse wave at only one position. It focusses on the analysis of the systolic blood pressure variations at the heart [20]. The idea is to estimate the increase in pressure ∆P at the heart, which occurs due to the superposition of ejecting blood into the aorta and pulse wave reflections, mostly caused by the aortic bifurcation [21].
In figure 3, an exemplary pulse wave with the pressure augmentation of ∆P is shown, with PP representing the total pulse pressure [22].
The augmentation index is defined as the relative pressure increase as given in equation 8.

$$AIx = \frac{\Delta P}{PP} \cdot 100\% \quad (8)$$

The instrumentation and the accompanying issues are very similar to the previously described problems, when analyzing the pulse wave velocity. Since the pressure pulse wave can typically not be measured directly at the ascending aorta, it is acquired at the extremities. This leads to an influence of the frequency dependent group delays on the signal morphologies and therefore distorts the measurement results.
In addition to this simplified discussion, several other scientific works have analyzed the unreliability of conventional pulse wave analysis technologies more detailed [23-25].

*B. Experimental determination of the aortic frequency response*

*1) Measurement Setup:*

As described before, significant measurement errors can be prevented by detecting the pulse wave directly at the aorta. To determine the aortic frequency response, the pulse wave has to be detected at the beginning as well as at the end of the aorta. For this, the bioimpedance plethysmography has proven to be a useful technique [26]. This approach is based on the assumption that the wavelengths composing the pulse wave signal are much higher than the length of the aorta. Therefore, neglecting occurring reflections is an acceptable simplification.
This measurement technique is based on the fact, that blood has a significantly higher conductivity than other tissue, for instance bones or fat [27]. By performing very sensitive bioimpedance measurements, the increase of blood volume in the area under observation, caused by the arriving pulse wave, can be detected. A compromise when using this technique is that only the pulse wave morphologies can be acquired, but not absolute pressure values.
The measurement setup used in this work is shown in figure 4. Two bioimpedance plethysmography measurements are performed simultaneously. To acquire the starting pulse wave, the surface electrodes (Ag/AgCl, Kendall H92SG, Medtronic, Minneapolis, MN, USA) of the first channel are positioned on the chest, above the aortic arch. After passing through the aorta, the arriving pulse wave is acquired at the groin (commonly iliac artery).
The utilized multi-channel bioimpedance measurement device was characterized in the past [26]. Its capability to acquire bioimpedances with signal-to-noise rations higher than 92 dB and therefore its applicability for high resolution pulse wave recordings was demonstrated. For better channel-separation, it is extended by additional decoupled current source modules [28]. The outer electrodes at the chest position are intended to apply an alternating current (AC) of 300 µA with a frequency of 77 kHz into the tissue in the field of the aortic arch. The inner electrodes are used to acquire the occurring voltage drop across the tissue. Their distance was adjusted to be about 4 cm. For the detection of the arriving pulse wave at the groin, the same voltage electrode distance is used, but a current of 1.5 mA with a frequency of 143 kHz is applied. Both the bioimpedances, $Z_{Chest}$ and $Z_{Leg}$ are measured simultaneously with a sampling rate of 1000 impedances per second.

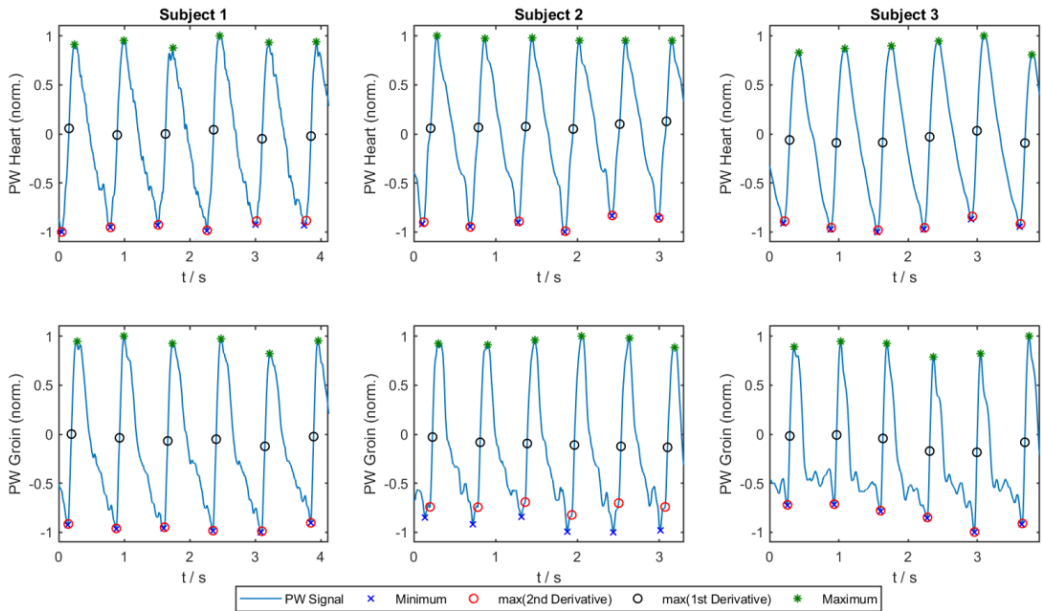

Fig. 7. Measurement results of the aortic pulse waves from 3 subjects. The upper three plots show the pulse wave signals acquired above the aortic arch. Below, the pulse waves measured at the groin are depicted. Additionally, the characteristic points to determine the pulse transit times are marked.

TABLE III
MEASURED PULSE TRANSIENT TIMES FOR THE FOUR COMMON DETECTION POINTS

|  | Subject 1 | Subject 2 | Subject 3 |
|---|---|---|---|
| $PTT_{foot}$ / ms | 91 | 29 | 46 |
| $PTT_{deri2}$ / ms | 88 | 80 | 38 |
| $PTT_{deri1}$ / ms | 38 | 18 | 2 |
| $PTT_{peak}$ / ms | 15 | 33 | -60 |

*2) Equivalent circuit of the aorta:*

Usually, linear time-invariant systems are characterized by applying a wide-band signal, such as an impulse or a step, to the system's input and measuring the output signal of the unknown system [26]. Obviously, it is not feasible to apply such an input signal to the aorta. Furthermore, the signal frequency range of the original input signal, the pulse wave, is very limited. Therefore, determining the transfer function via this typical procedure is not useful.

Instead, an electrical equivalent circuit is proposed, which models the behavior of the aorta and only the system parameters are determined by means of the pulse wave measurements. The equivalent circuit is based on other published works [19, 30, 31], which are focused on modelling the mechanical characteristics of the aorta as well. It emulates the dampening characteristic of the blood vessel as a second order RLC low-pass filter, as shown in figure 5.

In this model, the inductance $L_1$ represents the mass inertia of the blood volume, ejected by the heart pumping cycle. The capacity $C_1$ emulates the capability of the vessel to store blood volume by expending the vessel's volume when the pressure increases. This value is related to the arterial elasticity. When the pressure and thus the blood volume inside the aorta is increased, the blood flows into the peripheral arteries, slowed down by the flow resistance which is modelled as the load resistance $R_1$.

In table 2, the relevant electrical parameters and units are given, associated with the corresponding mechanical parameters and units [30].

The complex frequency response of the proposed RLC-model is shown in equation 9.

$$H_{LP2}(j\omega) = \frac{V_{out}(j\omega)}{V_{in}(j\omega)} = \frac{R_1}{R_1 + j\omega L_1 + (j\omega)^2 R_1 L_1 C_1} \quad (9)$$

As described above, the impedance plethysmography does only provide information about the pulse wave morphologies but not useful absolute pressure values. Therefore, the information about the absolute system gain cannot be measured. This leads to the fact that not all three variables can be determined, explicitly. Instead, only the relations between $L_1$, $R_1$ and $C_1$ can be analyzed when transposing the frequency response as shown in equation 10.

$$H_{LP2}(j\omega) = \frac{1}{1 + j\omega \frac{L_1}{R_1} + (j\omega)^2 L_1 C_1} \quad (10)$$

*3) Algorithm to determine the frequency response of the aorta:*

As described above, the mechanical behavior of the aorta is modelled as a second order low-pass filter. In equation 10, it can be seen that due to the transposing, and a substitution like in equation 11 yields a function of just two variables.

$$a_1 = \frac{L_1}{R_1}; \quad a_2 = L_1 C_1 \quad (11)$$

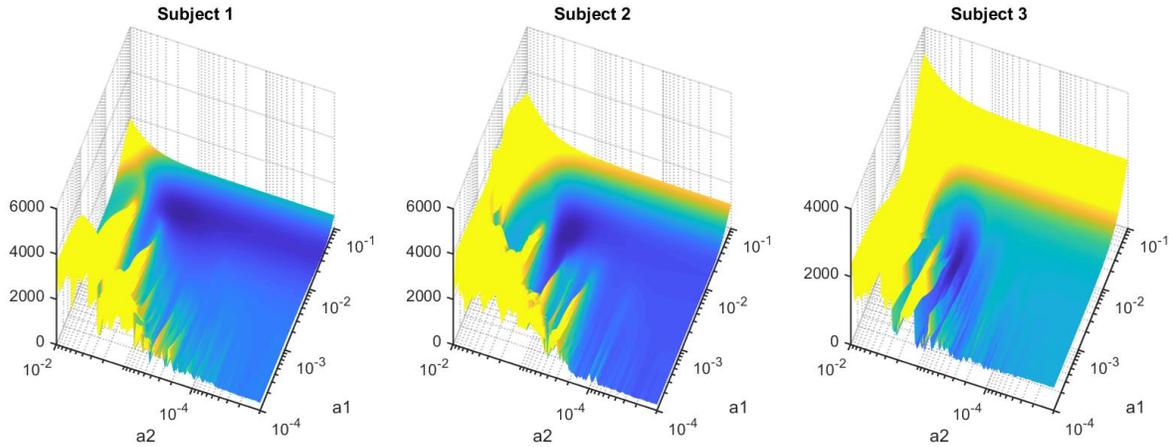

Fig. 8. Visualization of the error matrices to determine the optimum combinations of $a_1$ and $a_2$ for estimating the aortic frequency response of the three subjects. The dark blue regions represent the combinations, which lead to low errors.

To find the optimum combination of $a_1$ and $a_2$, which represents the actual aortic system best, an approximation algorithm is implemented, as shown in the block diagram in figure 6.

The input signal x(t), which corresponds to the measured pulse wave signal from the aortic arch, passes the actual aorta ($h_{true}(t)$) as well as an emulating low-pass filter ($h_{LP2}(t)$). Afterwards, both signals are normalized.

Using equation 12, the residual sum of squares of the output signals $y_{true}(t)$ and $y_{LP2}(t)$ is calculated in the comparison block. Afterwards, the determined error is stored in the error matrix $A_{error}(a_1,a_2)$. This procedure is repeated for realistic combinations of $a_1$ and $a_2$, which correspond with typical ranges of arterial cut-off frequencies between 1…16 Hz [32]. Finally, the position in $A_{error}(a_1, a_2)$ which contains the minimum error value is found, representing the optimum combination of $a_1$ and $a_2$.

$$A_{error}(a_1, a_2) = \int_{t_{min}}^{t_{max}} \left(y_{LP2}(t, a_1, a_2) - y_{true}(t)\right)^2 dt \quad (12)$$

### III. RESULTS AND DISCUSSION

The proposed measurement method is performed on three healthy young male subjects with approval of the ethics committee of the University of Lübeck.

Since respiration leads to geometrical changes within the thorax, this could also affect the acquired bioimpedance. To avoid this influence, the sitting subjects are asked to hold their breaths for about 10 seconds during the procedure.

Before beginning with the analysis of the pulse waves, both acquired signals from the chest and from the groin are digitally pre-processed. A first order low-pass filter with a cut-off frequency of $f_c$=15 Hz and a second order high-pass filter with a cut-off frequency $f_c$=0.3 Hz are applied to the signals. To avoid an influence on the signals' morphologies, caused by the non-constant group delay, zero phase filters are applied.

The acquired pulse wave signals are shown in figure 7. In the upper row, the signals from the electrodes positioned on the chests are plotted of all three subjects. Below, the corresponding simultaneously acquired signals from the groins are depicted. For better visualization, the signals are inverted and normalized.

In addition to the pulse waves, the corresponding characteristic points for common pulse wave velocity measurements are marked. Numerical methods are used to find the signals' minima (blue) and maxima (green). Furthermore, the maxima of the first (black) and second (red) derivatives are marked.

Using the marked characteristic points of the pulse waves, the common methods to estimate the PTT are applied. The results are shown in table 3.

Similar to the previously simulated signals and corresponding PTT values, these measurements show a high variability of the values depending on the subject and the used characteristic points, as well. Especially the results from subject 3 are very unrealistic. This might be caused by the specific pulse wave morphology. The determined $PTT_{peak}$ value is even negative, since the pulse wave maxima occur at the groin earlier than at the chest. This effect is quite realistic, assuming that occurring reflections, which reach the groin before the heart, increase the local pulse pressure. However, this measurement shows exemplary the difficulty of choosing useful characteristic pulse wave points to determine the transit time.

To determine the frequency responses of the aorta as described above, the optimum combination of $a_1$ and $a_2$ has to be found. Therefore, the minima of the calculated error matrices are detected. In figure 8, these error matrices are plotted, whereat the dark blue areas represent low errors and therefore appropriate combinations of $a_1$ and $a_2$. It can be seen that the minimum values are not salient, which might reduce the reliability of determining the optimum combination of $a_1$ and $a_2$.

The results of the proposed algorithm to estimate the aortic frequency responses are depicted in figure 9. The figure shows the amplitude responses, the phase responses and the group delays of the systems, which model the mechanical behavior

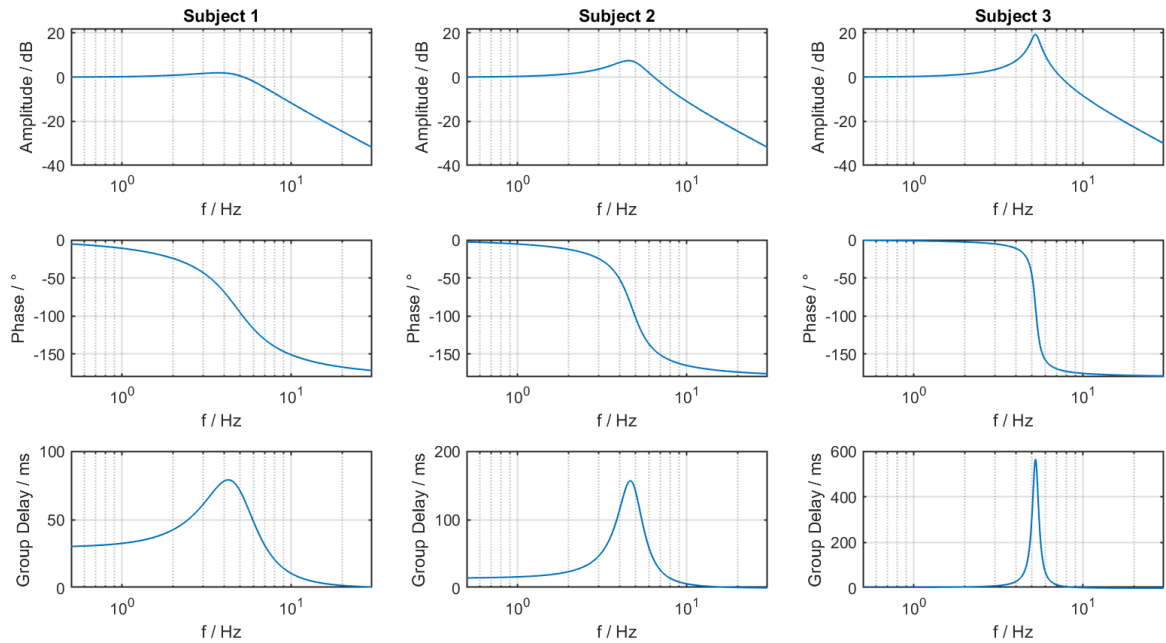

Fig. 9. Calculated aortic frequency responses and group delays of the thee subjects using the proposed modelling algorithm. In the bottom row, the corresponding group delays are plotted over frequency.

of the three subjects' aortas.

As predefined by the chosen model, all three systems have a low-pass behavior. The only differences are the cut-off frequencies and the heights of the amplitude overshoot. The resulting frequency responses are calculated to be as shown in equation 13-15, whereat the cut-off frequencies are $f_{c,Subject1}$=4.78 Hz, $f_{c,Subject2}$=4.76 Hz, $f_{c,Subject3}$=5.25 Hz.

$$H_{Subject1}(\omega) = \frac{1}{1 + j\omega \cdot 29.7 \cdot 10^{-3} + (j\omega)^2 \cdot 1.11 \cdot 10^{-3}} \quad (13)$$

$$H_{Subject2}(\omega) = \frac{1}{1 + j\omega \cdot 14.4 \cdot 10^{-3} + (j\omega)^2 \cdot 1.12 \cdot 10^{-3}} \quad (14)$$

$$H_{Subject3}(\omega) = \frac{1}{1 + j\omega \cdot 3.26 \cdot 10^{-3} + (j\omega)^2 \cdot 0.92 \cdot 10^{-3}} \quad (15)$$

The amplitude overshoot corresponds to the quality factor Q of the system and could be a more meaningful parameter, since it represents the system's sensitivity regarding ringing. It can be calculated with equation 16, whereat $f_0$ represents the resonance frequency and B represents the bandwidth of the system.

$$Q = \frac{f_0}{B} \quad (16)$$

The resulting quality factors are $Q_1$=1.9, $Q_2$=2.6, $Q_3$=5.9, indicating that the aorta model related to subject 3 is more prone to ringing than the others.

In the bottom plots of figure 9, the frequency dependency of the group delays can be seen, which means the actual pulse transit times over frequency. The strong variations within the frequency range of typical pulse wave signals explain the significant morphology changes of the pulse wave when passing the aorta and therefore the issue of common pulse wave analysis methods.

These results have to be considered carefully, since the plots look very different, even though the physical conditions of the three subjects are very similar.

Finally, to compare the actual behavior of the aorta with the modelled systems, the bioimpedance signal from the groin (red) is plotted in combination with the model output signal (blue) in figure 10. Additionally, the bioimpedance signal from the chest (black) is plotted.

In the first plot it can be seen, that the output of the model has a similar morphology as the actual output of the aorta. Especially, the time delay between input and output signal, seems to be quite realistic, when focusing on the significant rising slopes.

This time delay cannot be seen that clearly in the second plot. Instead, the delays of the pulse wave minima seem to be modelled very realistically.

In the third plot, the model output is oscillating, caused by the high quality factor of the system. This does not represent the actual output of the aorta and could be an issue of the applied model.

The measurements pointed out that the approach of modelling the aorta as an RLC low-pass leads to realistic results for the three exemplary subjects. However, comparing the output signals of the actual aorta and the modelled aorta shows that there are still significant differences in the morphologies. It is conceivable that a higher system order of the model could yield more suitable output signals.

It has to be considered that the bioimpedance plethysmography via surface electrodes uses an electrical field, whose geometrical distribution cannot exactly be determined. This fact complicates the measurement of the distance between both the sensing positions.

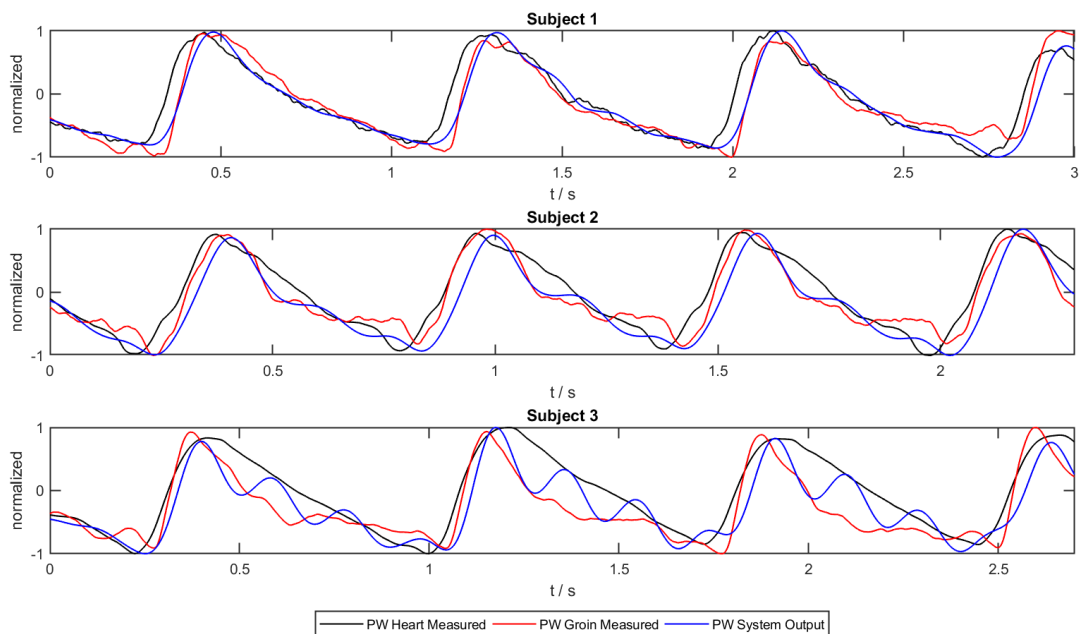

Fig. 10. Comparison of the measured input signal of the aorta and the resulting actual and modelled output signals.

Obviously, the measurement results of only three subjects in this study are not sufficient to judge if the proposed approach is very accurate and can therefore become a new standard. However, these preliminary results have proven that the determination of the aortic frequency response by performing two simultaneous bioimpedance measurement at the chest and at the groin is a reasonable measurement approach to reduce the issues which occur in common measurement techniques.

Not only the proposed positioning of the sensors to acquire the aortic pulse wave instead of the peripheral wave is beneficial. Especially, the processing of the whole pulse wave morphologies for estimating the physical behavior of the aorta eliminates the problems regarding the selection of characteristic points to calculate the PWV [33-35]. Nevertheless, assuming the aorta to be an ideal linear system yields obviously to limitations, as well.

## IV. CONCLUSION

Common techniques to determine the mechanical behavior of the aorta via pulse wave analysis are sensitive to changes of the specific measurement conditions. Minor changes of the heart rate or wrong assumptions about the aorta geometry lead to significant measurement errors. In real measurement scenarios, these occurring errors can even be higher than the precision which is mandatory to diagnose the state of the aorta.

The proposed measurement setup to detect the pulse wave at the aortic arch and at the groin by performing simultaneous bioimpedance measurements seems to be more reliable. Since both, the input and the output of the unknown system are measured directly, there are no complex algorithms required to estimate the pulse wave morphologies.

Modelling the aorta as an electrical RLC circuit has been proven to be a simple and useful approach. A major advantage is that not only a single point of the pulse wave is analyzed, like in many pulse transient time based approaches, but the information of the whole pulse wave signal. However, in the proposed measurements it could also be seen that this model does not work well for all subjects. The usability of more complex models has to be analyzed in the future. In addition to linear time-invariant systems, these models could also be non-linear systems. Furthermore, the occurring reflections within the blood vessels could be included in a model. Therefore, the transmission line theory could be a promising approach.

Since the proposed first measurements represent only three subjects and no ground truth was provided, more extensive studies on subjects are necessary to analyze and compare the medical usefulness.

The authors declare that they have no conflict of interest.

## REFERENCES

[1] G. A. Roth *et al*., "Global and Regional Patterns in Cardiovascular Mortality From 1990 to 2013," *Circulation*, vol. 132, no. 17, pp. 1667–1678, 2015.
[2] F. U. Mattace-Raso *et al*., "Arterial Stiffness and Risk of Coronary Heart Disease and Stroke," *Circulation*, vol. 113, no. 5, pp. 657–663, Jul. 2006.
[3] A. P. Avolio *et al*., "Arterial blood pressure measurement and pulse wave analysis--their role in enhancing cardiovascular assessment," *Physiological Measurement*, vol. 31, no. 1, 2009.
[4] S. Wassertheurer *et al*., "A new oscillometric method for pulse wave analysis: comparison with a common tonometric method," *Journal of Human Hypertension*, vol. 24, no. 8, pp. 498–504, 2010.
[5] L. Luzardo *et al*., "24 Hour Ambulatory Recording Of Aortic Pulse Wave Velocity And Central Systolic Augmentation—A Feasibility Study," *Journal of Hypertension*, vol. 29, 2011.

[6] M. W. Rajzer *et al*., "Comparison of aortic pulse wave velocity measured by three techniques: Complior, SphygmoCor and Arteriograph," *Journal of Hypertension*, vol. 26, no. 10, pp. 2001–2007, 2008.
[7] A. P. Guerin *et al*., "Impact of Aortic Stiffness Attenuation on Survival of Patients in End-Stage Renal Failure," *Circulation*, vol. 103, no. 7, pp. 987–992, 2001.
[8] S. Laurent *et al*., "Expert consensus document on arterial stiffness: methodological issues and clinical applications," *European Heart Journal*, vol. 27, no. 21, pp. 2588–2605, 2006.
[9] S. Laurent *et al*., "Aortic Stiffness Is an Independent Predictor of All-Cause and Cardiovascular Mortality in Hypertensive Patients," *Hypertension*, vol. 37, no. 5, pp. 1236–1241, 2001.
[10] L. M. V. Bortel *et al*., "Expert consensus document on the measurement of aortic stiffness in daily practice using carotid-femoral pulse wave velocity," *Journal of Hypertension*, vol. 30, no. 3, pp. 445–448, 2012.
[11] J. Sugawara *et al*., "Brachial-ankle pulse wave velocity: an index of central arterial stiffness?," *J Hum Hypertens*, vol. 19, no. 5, pp. 401–406, 2005.
[12] H. Elkenani *et al*., "Numerical Investigation of Pulse Wave Propagation in Arteries Using Fluid Structure Interaction Capabilities," *Computational and Mathematical Methods in Medicine*, vol. 2017, pp. 1–7, 2017.
[13] D. J. Korteweg, "Ueber die Fortpflanzungsgeschwindigkeit des Schalles in elastischen Röhren," *Annalen der Physik und Chemie*, vol. 241, no. 12, pp. 525–542, 1878.
[14] F. U. S. Mattace-Raso *et al*., "Determinants of pulse wave velocity in healthy people and in the presence of cardiovascular risk factors: 'establishing normal and reference values,'" *European Heart Journal*, vol. 31, no. 19, pp. 2338–2350, Jul. 2010.
[15] B. E. Westerhof *et al*., "Location of a Reflection Site Is Elusive," *Hypertension*, vol. 52, no. 3, pp. 478–483, 2008.
[16] J. Blacher, "Aortic pulse wave velocity (PWV): Marker of cardiovascular risk in hypertensive patients," *American Journal of Hypertension*, vol. 11, no. 4, 1998.
[17] Y. Choi *et al*., "Noninvasive cuffless blood pressure estimation using pulse transit time and Hilbert–Huang transform," *Computers & Electrical Engineering*, vol. 39, no. 1, pp. 103–111, 2013.
[18] P. Boutouyrie *et al*., "Assessment of pulse wave velocity," *Artery Research*, vol. 3, no. 1, pp. 3-8, 2009.
[19] T. S. Manning *et al*., "Validity and Reliability of Diastolic Pulse Contour Analysis (Windkessel Model) in Humans," *Hypertension*, vol. 39, no. 5, pp. 963–968, 2002.
[20] I. B. Wilkinson *et al*., "Reproducibility of pulse wave velocity and augmentation index measured by pulse wave analysis," *Journal of Hypertension*, vol. 16, no. 12, pp. 2079–2084, 1998.
[21] Yasmin, and M. J. Brown, "Similarities and differences between augmentation index and pulse wave velocity in the assessment of arterial stiffness," *Q J Med*, vol. 92, no. 10, pp. 595–600, Jan. 1999.
[22] I. B. Wilkinson *et al*., "Increased central pulse pressure and augmentation index in subjects with hypercholesterolemia," *Journal of the American College of Cardiology*, vol. 39, no. 6, pp. 1005–1011, 2002.
[23] M. Willemet *et al*., "A database of virtual healthy subjects to assess the accuracy of foot-to-foot pulse wave velocities for estimation of aortic stiffness," *Am J Physiol Heart Circ Physiol,* vol. 309, pp. H663-H675, 2015.
[24] B. Trachet *et* al., "Numerical Validation of a New Method to Assess Aortic Pulse Wave Velocity from a Single Recording of a Brachial Artery Waveform with an Occluding Cuff," *Annals of Biomedical Engineering,* vol. 38, no. 3, pp. 876-888, 2010.
[25] J. Davies and A. Struthers, "Pulse wave analysis and pulse wave velocity: a critical review of their strengths and weaknesses," *Journal of Hypertension*, vol. 21, no. 3, pp. 463-472, 2003.
[26] R. Kusche *et al*., "A Multichannel Real-Time Bioimpedance Measurement Device for Pulse Wave Analysis," *IEEE Transactions on Biomedical Circuits and Systems*, vol. 12, no. 3, pp. 614–622, 2018.
[27] S. Grimnes, and O. G. Martinsen, *Bioelectricity and Bioimpedance Basics* (2nd ed.), New York, Academic Press, 2008.
[28] R. Kusche *et al*., "Galvanically Decoupled Current Source Modules for Multi-Channel Bioimpedance Measurement Systems," *Electronics*, vol. 6, no. 4, p. 90, 2017.
[29] J. G. Proakis and D. Manolakis, Digital Signal Processing: principles, algorithms, and applications (3rd ed.), Prentice-Hall, Inc. Upper Saddle River, NJ.
[30] T. Pustelny *et al*., "Design and numerical analyses of the human greater circulatory system," *The European Physical Journal Special Topics*, vol. 154, no. 1, pp. 171–174, 2008.
[31] W. Welkowitz *et al*., "Noninvasive estimation of cardiac output," *IEEE Transactions on Biomedical Engineering*, vol. 38, no. 11, pp. 1100–1105, 1991.
[32] A. P. Avolio, "Multi-branched model of the human arterial system," *Med. & Biol. Eng. & Comput*, vol 18, no. 6, pp. 709-718, 1980.
[33] N. R. Gaddum *et al*., „A technical assessment of pulse wave velocity algorithms applied to non-invasive arterial waveforms," Ann Biomed Eng, vol. 41, no. 12, pp. 2617-2629, 2013.
[34] O. Vardoulis *et al*., "On the estimation of total arterial compliance from aortic pulse wave velocity," Ann Biomed Eng, vol. 40, no. 12, pp. 2619-2626, 2012.
[35] E. Hermeling *et al*., "The dicrotic notch as alternative time-reference point to measure local pulse wave velocity in the carotid artery by means of ultrasonography," J Hypertens, vol. 27, no. 10, pp. 2028-2035.